\documentclass[twocolumn]{IEEEtran} 
\def\BibTeX{{\rm B\kern-.05em{\sc i\kern-.025em b}\kern-.08em
    T\kern-.1667em\lower.7ex\hbox{E}\kern-.125emX}}

\RequirePackage{times}[1995/08/06]   
\usepackage{graphicx}

\newcommand{\degree}{~$^{\circ}\mbox{C}\;$}

\begin{document}

\title{Recent Progress in CdTe and CdZnTe  Detectors }

\author{%
\thanks{Manuscript received Febrary 2, 2001.}
Tadayuki Takahashi 
\thanks{T. Takahashi is
with the Institute of Space and Astronautical Science (ISAS),
Sagamihara, Kanagawa 229-8510, Japan and is also with
Department of Physics, the University of Tokyo, Bunkyo, Tokyo
113-0033, Japan. (telephone: 81-42-759-8150, e-mail:
takahashi@astro.isas.ac.jp)} 
and Shin Watanabe
\thanks{S. Watanabe is 
with the Institute of Space and Astronautical Science (ISAS),
Sagamihara, Kanagawa 229-8510, Japan and is also with
Department of Physics, the University of Tokyo, Bunkyo, Tokyo
113-0033, Japan.(telephone: 81-42-759-8151, e-mail:
watanabe@astro.isas.ac.jp)}}

\markboth{IEEE Transactions On Nuclear Science, Vol. XX, No. Y, Month
2000}{Takahashi and Watanabe: Recent Progress in CdTe and CdZnTe}

\maketitle

\begin{abstract}

Cadmium telluride (CdTe) and cadmium zinc telluride (CdZnTe) have been
regarded as promising semiconductor materials for hard X-ray and
$\gamma$-ray detection. The high atomic number of the materials
(Z$_{Cd}$ =48, Z$_{Te}$=52) gives a high quantum efficiency in
comparison with Si.  The large band-gap energy (Eg $\sim$ 1.5 eV)
allows us to operate the detector at room temperature.  However, a
considerable amount of charge loss in these detectors produces a
reduced energy resolution. This problem arises due to the low mobility
and short lifetime of holes. Recently, significant improvements have
been achieved to improve the spectral properties based on the advances
in the production of crystals and in the design of electrodes.  In
this overview talk, we summarize (1) advantages and disadvantages of
CdTe and CdZnTe semiconductor detectors and (2) technique for
improving  energy resolution and photopeak efficiencies.
Applications of these imaging detectors in future hard X-ray and
gamma-ray astronomy missions are briefly discussed.

\end{abstract}


\section{Introduction}

\PARstart{T}{here} are increasing demands for new semiconductor detectors 
capable of detecting hard X-ray and $\gamma$-rays. For imaging
devices, their good energy resolution and the ability to fabricate
compact arrays are very attractive features in comparison with
inorganic scintillation detectors coupled to either
photodiodes or photomultiplier tubes.

Despite the
 excellent energy resolution and
charge-transport properties of silicon (Si) and germanium (Ge) detectors, 
their low  stopping power for high energy photons limits their
application to hard X-ray and $\gamma$-ray detection.
Furthermore, the small band gap of germanium forces us to
 operate the
detector at cryogenic temperatures.
Therefore, room-temperature semiconductors with high atomic numbers and
wide band gaps have long been under development. 
These materials are useful not only in medical and industrial
imaging systems but also in  detectors for high energy particle-
and astrophysics.

Cadmium Telluride (CdTe) has been regarded as a promising
semiconductor material for hard X-ray and $\gamma$-ray detection since
the early 1970's. The high atomic number of the materials gives a high
quantum efficiency suitable for a detector operating typically in the
10$-$500 keV range. A large band-gap energy ($E_{gap}$ = 1.44 eV)
allows us to operate these detectors at room temperature.  However, it
became clear that it would be much more difficult to use CdTe in
comparison with Si and Ge, especially for nuclear spectroscopy, where
good spectroscopic performance is desired.  The lack of stability also
limited the usefulness of CdTe detectors in the past.  Problems and
efforts to improve the properties of CdTe detectors since its
beginning are available in review
articles\cite{ref:Richter,ref:Siffert}.

In the 1990's, the remarkable progress in the technology of producing
a high quality single crystal of CdTe and the emergence of Cadmium
Zinc Telluride (CdZnTe) have dramatically changed the situations of
high resolution room temperature detectors
\cite{ref:Eisen}. Furthermore, advances in application specific
integrated circuits (ASICs) leads to fabrication of position sensitive
detectors in the form of strip or pixellated
detectors\cite{ref:Augustine,ref:Weilhammer}.  Here we summarize
the recent progress of the development of CdTe and CdZnTe detectors.

\section{CdTe and  CdZnTe  semiconductor}

\begin{figure}
\label{figure:first}
\vspace{1mm}
\centerline{\includegraphics[width=3.15in,clip]{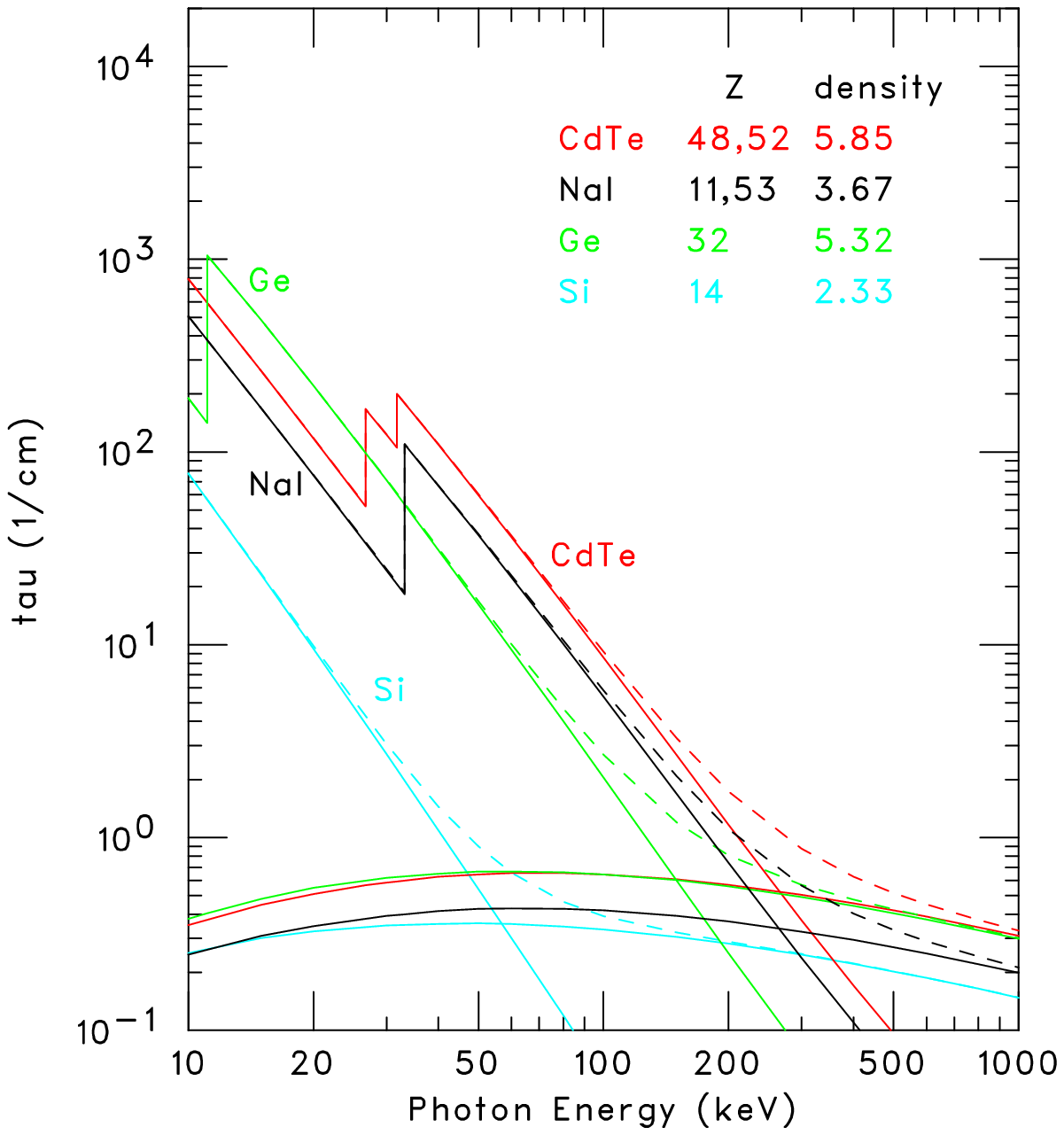}}
\caption{Linear Attenuation Coefficients for photo absorption and Compton
scattering in CdTe, Si, Ge,  and NaI($Tl$)}
\end{figure}

\begin{table}[htb]
\label{tab:Characteristics}
\caption{Properties of the semiconductors}
\begin{center}
\begin{tabular}{llllll}
\hline
semi-     & density         & Z & $E_{\rm gap}$  & $\epsilon$ & $X_{0}$\\
conductor &${\rm [g/cm^3]}$&   & [eV]       & [eV]  & [cm] \\
\hline
Si           & 2.33 &14    & 1.12 & 3.6 & 9.37\\
Ge           & 5.33 &32    & 0.67 & 2.9 &2.30\\
CdTe         & 5.85 &48,52 & 1.44 & 4.43 &1.52\\
CdZnTe       & 5.81 & & 1.6 &4.6 & \\
${\rm HgI_2}$& 6.40 &80,53 & 2.13 & 4.2  & 1.16\\
${\rm GaAs}$& 5.32 &31, 33 & 1.42 & 4.3& 2.29 \\
\hline
\end{tabular}
\end{center}

\hspace{1cm}
\begin{tabular}{lcl} 
$E_{\rm gap}$ &:& band gap energy \\
$\epsilon$ &:& an ionization potential \\
$X_{0}$ &:& radiation length \\
\end{tabular}
\end{table}

\begin{figure}
\label{figure:second}
\vspace{1mm}
\centerline{\includegraphics[width=3.15in,clip]{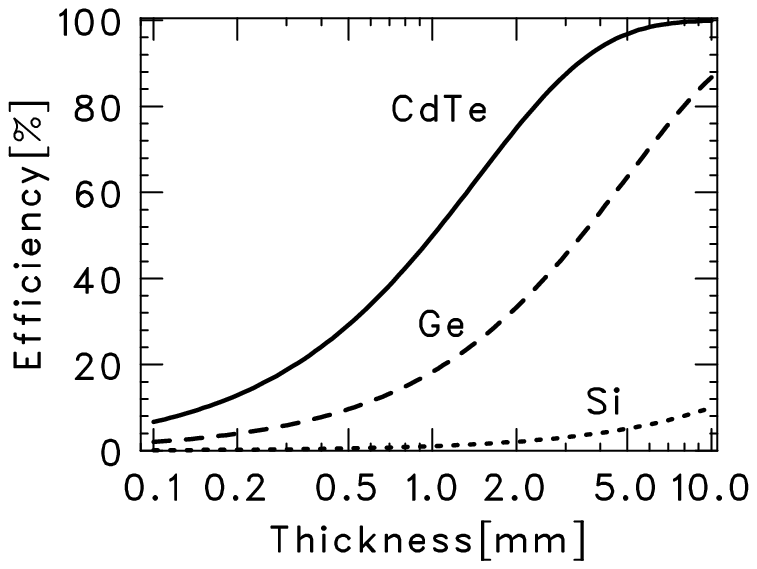}}
\caption{Detection efficiency for 100 keV $\gamma$-ray photon in various thickness
of CdTe, Si and Ge}
\end{figure}

\begin{figure}
\label{figure:third}
\vspace{1mm}
\centerline{\includegraphics[width=3.15in,clip]{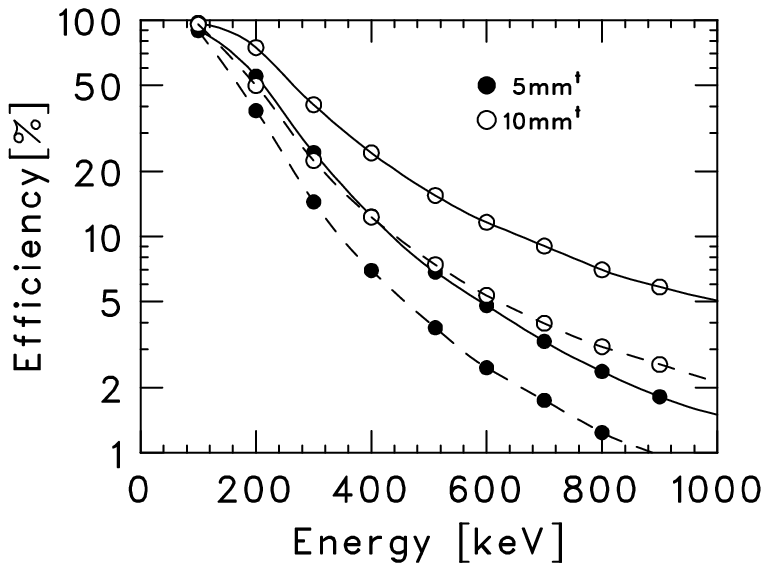}}
\caption{Detection efficiency of CdTe at different energies for a
thickness of 5 mm (the closed cicles) and 10 mm (the open circles).
Dashed line : Photoelectric absorption only. Solid line : Efficiencies
in which multiple Compton scatterings are taken into account. In the
simulation, collimated $\gamma$-rays are irradiated at the center of a
detector with a surface area of 1 cm $\times$ 1 cm}
\end{figure}

Table~I~ shows the physical characteristics of the elemental and
compound semiconductors. Among the range of semiconductor detectors
available for $\gamma$-ray detection, CdTe (and CdZnTe) have a
privileged position\cite{ref:Scheiber,ref:Iwanczyk,ref:Bencivelli},
because of their high density and the high atomic number of their
components, as well as a wide bandgap.  As shown in Fig.~1~, CdTe has
a high photoelectric attenuation coefficient.
Photoelectric absorption is the main process up to 300 keV for CdTe,
as compared to 60 keV for Si and 150 keV for Ge.  Fig.~2~ shows the
efficiency for 100 keV $\gamma$-rays.  Even a detector with a
thickness of 0.5 mm provides a good detection efficiency for
$\gamma$-rays. Efficiencies calculated with a simulation code
\cite{ref:Geant4} are shown in Fig.~3~ for a detector with a thickness
of 5 mm and 10 mm.  For a 511 keV photon, an absorption efficiency of
15.5\% can be obtained for a 10 mm cube detector. These efficiencies,
however, assume that $\gamma$-ray photons deposit all their energy in
the detector volume and that we can collect all electron-hole pairs
generated in the detector. It has been pointed out that the
considerable amount of charge loss in CdTe and CdZnTe limits their
capability as high resolution spectrometers\cite{ref:Siffert}. As
shown later, this problem is due mainly to the poor charge transport
properties, especially for holes.  Incomplete charge collection could
limit the thickness and, thus, the volume of detectors which in turn
limits the usefulness of the detector.

\subsection{HPB-grown CdZnTe}

Very recently, Cd$_{1-x}$Zn$_{x}$Te grown by the High Pressure
Bridgman (HPB) technique has emerged as a new choice for room
temperature detectors\cite{ref:Doty,ref:Butler1,ref:Szeles1}. By
increasing theZn concentration, the bandgap of the material is
increased from 1.44 eV (for CdTe) to 2.2 eV (for
ZnTe)\cite{ref:Toney}. With the most widely used composition of $x$ =
0.08 $-$ 0.15, CdZnTe has a bandgap of $\sim$ 1.6 eV and displays
high electrical resistivity $\rho =$ (1.0 $-$ 4.0) $\times$ 10$^{10}$
$\Omega$ cm (n-type), which is close to the theoretical maximum
allowed by the bandgap\cite{ref:Szeles1}.  As shown in the current
voltage (I-V) characteristics (Fig.~4~), the leakage current of a
4$\times$4$\times$2 mm$^{3}$ CdZnTe detector is 4 nA at
20~\degree. The low leakage current due to the high resistivity of
HPB-grown CdZnTe and its good electron transport of $\mu_{e} \tau_{e}$
= (0.5 $-$ 5.0) $\times 10^{-3}$ cm$^{2}$/V, where $\mu\tau$ is
mobility-lifetime product, results in good spectral performance when
operated at room temperature. These advantages over conventional CdTe
have lead to the early success of small volume single-element planar
detectors and simple multi-electrode CdZnTe devices have stimulated
more advanced application in a variety of
fields\cite{ref:Szeles2,ref:Eisen2}.  Nevertheless, there are a few
drawbacks associated with this new detector material. Although the
ingots are of very large volume, the present HPB technique yields only
polycrystals with a non-uniform distribution of $\mu\tau$ products.
The IR transmission image of the large wafer shows grain boundaries
and a distribution of Te-decorated features
\cite{ref:Stahle}. Extensive studies indicate that the grain
boundaries decorated with Te inclusions or twin boundaries have an
adverse effect on carrier transport in CdZnTe and result in poor
spectral response\cite{ref:Szeles2,ref:Stahle,ref:Burger}. At present,
the yield of HPB CdZnTe dies suitable for the fabrication of
large-area ($>$ 10 $\times$ 10 mm$^{2}$) X- and $\gamma$-ray imaging
devices is very low \cite{ref:Szeles2}. Additionally, the $\mu\tau$
products of holes in CdZnTe detectors is $\sim$2$\times$10$^{-5}$
cm$^{2}$/V and is almost one order of magnitude smaller than that of
the recent CdTe detector. This is probably due to some degradation in
properties introduced by adding Zn.

\begin{figure}
\label{figure:4th}
\vspace{1mm}
\centerline{\includegraphics[width=3.15in,clip]{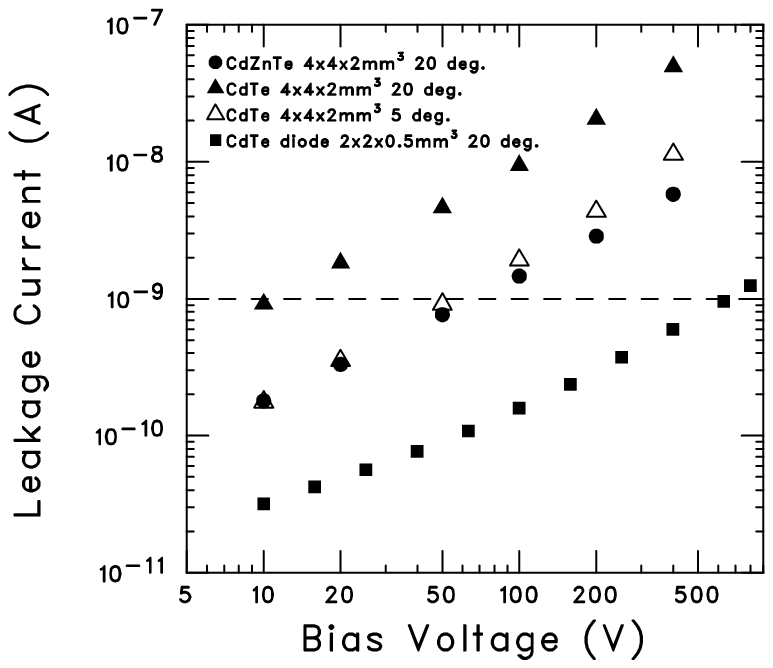}}
\caption[]{Current voltage (I-V) characteristics at 20~\degree
of the CdZnTe and CdTe detectors with dimensions of 4$\times$4 mm$^{2}$ and a
2 mm thickness. The leakage current  for the
CdTe detector measured at 5~\degree  and
the CdTe diode (described later) at 20~\degree. The dimensions of the CdTe
diode is 2 $\times$ 2 $\times$ 0.5
mm$^{3}$   are shown for comparison. The CdZnTe
detector was manufactured by eV Products and the CdTe detector
was manufactured by ACRORAD.}
\end{figure}

\subsection{THM-grown CdTe}
After continuous efforts for more than two decades to improve the
performance of CdTe crystals, the technique for growing a large single
CdTe crystal with good charge transport properties seems to be
established\cite{ref:Funaki} CdTe is usually grown by the Traveling
Heater Method (THM).  After careful thermal treatment and the
selection of proper crystal orientation for the electrode system, it
is reported that the CdTe wafer displays good charge transport
properties for both electrons ($\mu_{e} \tau_{e}$ = 1$-$2 $\times
10^{-3}$ cm$^{2}$/V ) and holes ($\mu_{h} \tau_{h}$ = 1 $\times
10^{-4}$ cm$^{2}$/V ).  The electrical resistivity of
$\sim$1$\times$10$^{9}$ $\Omega$ cm (p-type) is achieved by
compensating the native defects with Cl. Furthermore, with the usual
electrode configuration with Pt which forms Ohmic contacts, the
detector is free from problems with the stability\cite{ref:Funaki}.  There
are no grain boundaries nor a distribution of Te inclusion in a wafer,
since it is a single crystal\cite{ref:Stahle}. The uniform charge
transport properties of the wafer are very important aspect not only
for fabricating large area strip or pixel detectors but also for
constructing a large scale $\gamma$-ray camera with many individual
detectors.  Single crystals of 50 mm diameter are now commercially
available for THM-CdTe. The grown crystal is large enough to obtain (1
1 1)- oriented single crystal wafers with an area as large as
30$\times$ 30 mm$^{2}$.  A similar approach to growing high quality
CdTe single crystals is reported with THM \cite{ref:Fougeres} and in a
conventional vertical Bridgman furnace\cite{ref:Szeles2}.

\begin{figure}
\label{figure:5th}
\vspace{1mm}
\centerline{\includegraphics[width=3.15in,clip]{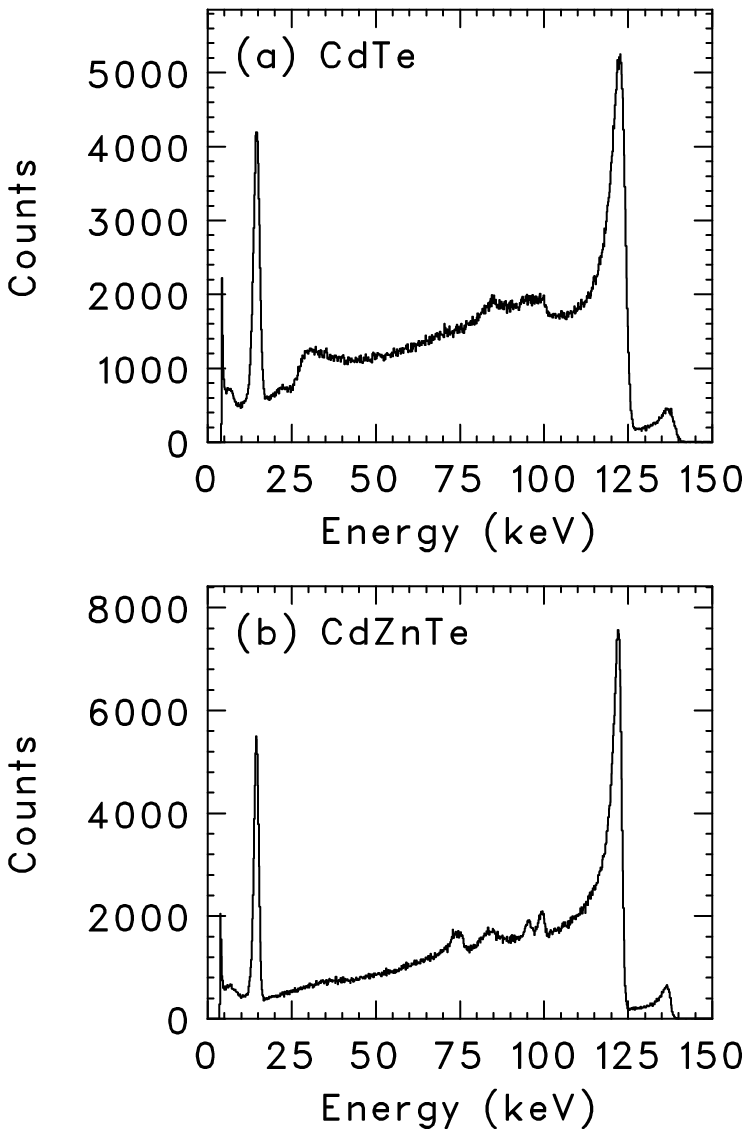}}
\caption{$^{57}$Co spectra obtained with (a) CdTe and (b) CdZnTe detectors
at 20~\degree obtained from the same detectors used in Fig.~4~. The 
applied bias voltage is 100 V and 300 V for CdTe and CdZnTe,
respectively.
The shapeing time was set at 0.5 $\mu$s.  }
\end{figure}

\subsection{Energy spectra of CdTe and CdZnTe}

 Because of the very high resistivity, CdTe and CdZnTe are regarded as
semi-insulating materials and usually operated as a ``solid ionization
chamber''\cite{ref:Knoll}. The electron-hole pairs generated in the
detector are collected by applying an appropriate bias voltage.
Figs.~5~ (a) and (b) show the energy spectra of $\gamma$-rays from
$^{57}$Co obtained at 20~\degree with the CdTe and CdZnTe detectors
used in the measurement of the I-V curve (Fig.~5~). In the
measurement, the charge signal is integrated in the Clear Pulse
CP-5102 CSA and shaped by an ORTEC 571 amplifier. The time constant of
the shaping amplifier was set at 0.5 $\mu$s for both detectors.  In
order to minimize the effect of the incompleteness of the hole
collection, we irradiated $\gamma$-rays from radioactive sources on
the negative electrode (cathode). The applied bias voltages are chosen
to be 100 V for the CdTe and 300 V for the CdZnTe detector so that we
obtain the comparable leakage current.  The energy resolution (FWHM)
of the 122 keV line is 6.9 keV for the CdTe detector and 4.2 keV for
CdZnTe.

\section{Effects of low transport of holes}

 Collecting full
information due to the transit of both electrons and holes is important
for obtaining the ultimate energy resolution from the device.  The mean drift
path of the charge carrier is expressed as the product of $\mu\tau$
and $E$, where $E$ is the applied electric field in the device.
Due to the slow mobility and
short lifetime of holes, the thickness of the detector should be
smaller than $\mu_{h}\tau_{h} E$, where $\mu_{h}$ and $\tau_{h}$ are
the mobility and lifetime of holes. If a
detector with a thickness $l > \mu_{h}\tau_{h} E$ is used, only a
fraction of the generated signal charge is induced at the detector
electrode. The fraction and the
resultant pulse height  depend on the interaction depth.  This
position dependency produces a shoulder (tailing) in the peaks of
$\gamma$-ray lines towards the low energy region, which is seen in
$^{57}$Co spectra with CdTe and CdZnTe shown in Fig.~5~. Due
to a distortion of the spectrum. The energy resolution does not
reach the theoretical limit expected from statistical fluctuations in
the  number of  electron-hole pairs and the Fano factor.

The pulse height as a function of the depth of interaction for a simple
planar detector is
given by the Hecht equation\cite{ref:Hecht}:

\begin{eqnarray}
PH(z) &\propto& 
n_{0} (\mu_{e}\tau_{e} E (1- \exp(-\frac{d-z}{\mu_{e}\tau_{e}E}))\\
& & + \mu_{h}\tau_{h} E (1- \exp(-\frac{z}{\mu_{h}\tau_{h}E})) )
\end{eqnarray}
where $n_{0}$ is the number of electron-hole pairs generated in the
detector and $d$ is the detector thickness. The depth, $z$ is measured
from the cathode. The reduction of the pulse height is severe for the
event which takes place close to the anode electrode of a 2 mm thick
detector, where the signal is mostly due to the hole transit. Only
16\% of the signal is collected for $\mu_{h}\tau_{h}$ =
2$\times$10$^{-5}$ cm$^{2}$/V (typical number for the recent HPB-grown
CdZnTe detectors) as compared to the case for $\mu_{h}\tau_{h} =
\infty$ with a bias voltage of 300 V. This effect is clearly seen
in the $^{241}$Am spectrum shown in Fig.~6~ 
which was obtained with a CdZnTe detector
with a thickness of 2 mm. When the detector is irradiated by
$\gamma$-rays from the cathode face, $\gamma$-ray peaks are clearly
resolved. The energy resolution at 60 keV is  3.3 keV. However,
if we irradiate $\gamma$-rays from the anode face, the peak structure
is smeared out completely. The situation is slightly better for CdTe
which has higher $\mu_{h}\tau_{h}$, but the spectral performance is
still limited.

\begin{figure}
\label{figure:6th}
\vspace{1mm}
\centerline{\includegraphics[width=3.15in,clip]{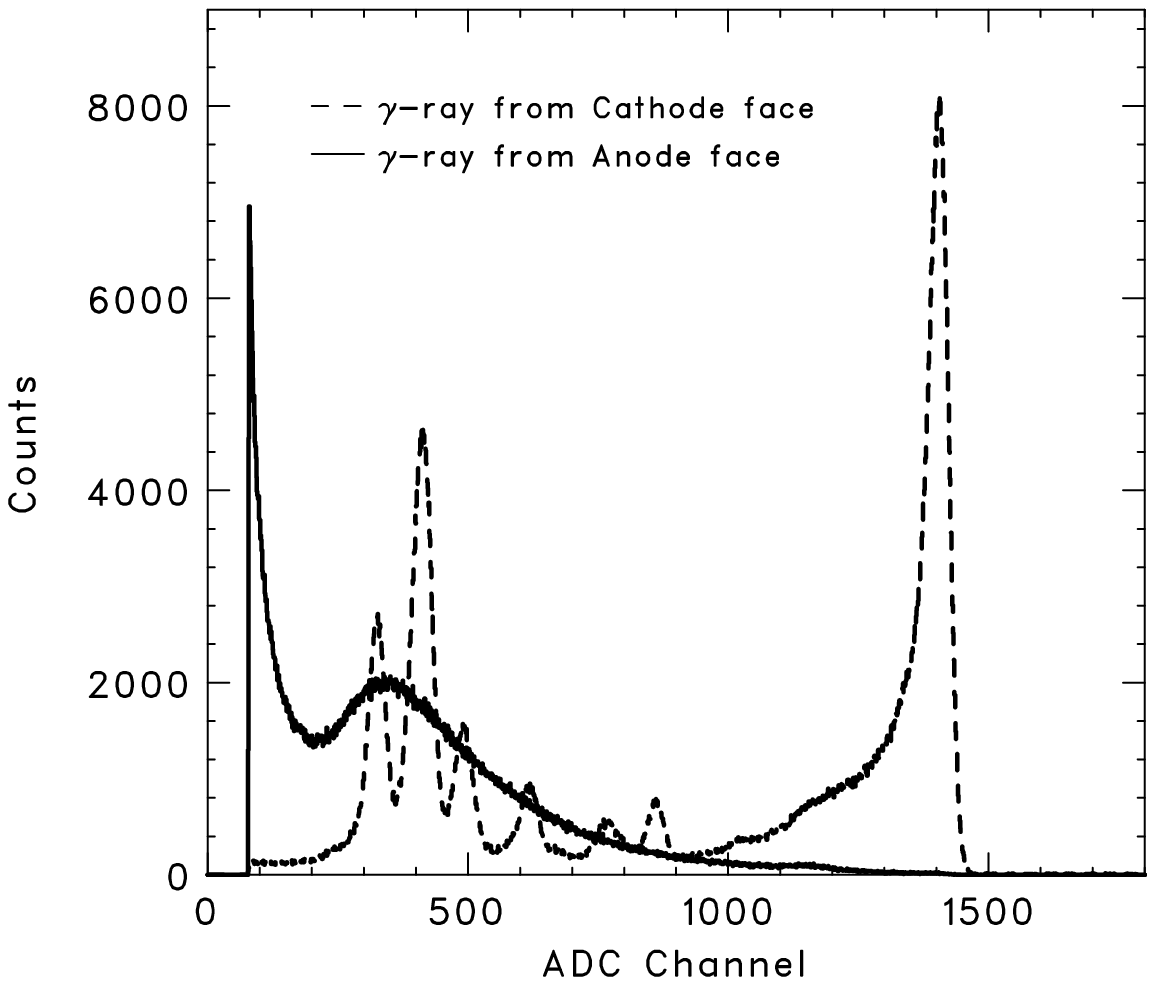}}
\caption{$^{241}$Am spectra obtained with the 2 mm thick CdZnTe
detector used in Fig.~4~.  The applied bias voltage is 300 V and the
operating temperature is 5~\degree.  $\gamma$-rays are irradiated from
the cathode side (solid line) and the anode side (dashed line).}
\end{figure}

\begin{figure}
\label{figure:7th}
\vspace{1mm}
\centerline{\includegraphics[width=3.15in,clip]{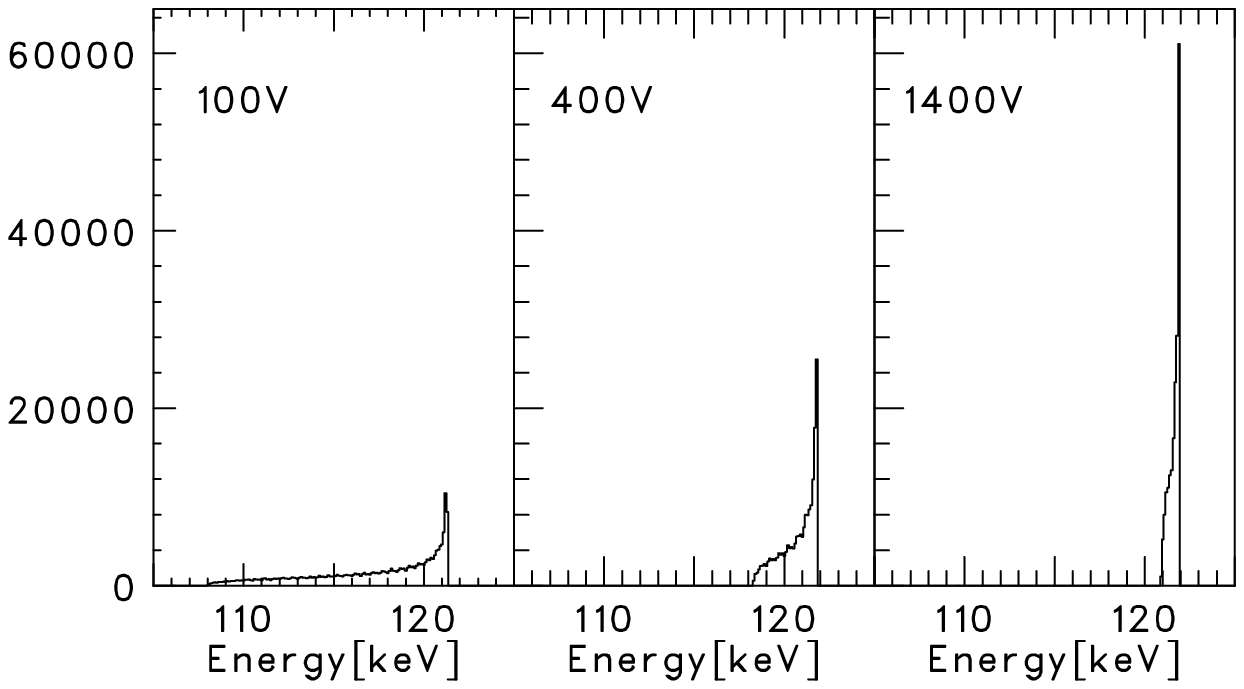}}
\caption{Simulated spectra of the 122 keV line of $^{57}$Co obtained
with (a) 100V, (b) 400 V, and (c) 1400 V. The detector thickness is
0.5 mm.  The low energy tail in the spectrum is solely due to the hole
trapping. For the this thickness, 1400 V is required for eliminating
low energy tail and obtaining an energy resolution close to the
theretical limit. The assumed $\mu_{e}\tau_{e}$ is 2$\times$10$^{-3}$
cm$^{2}$/V and $\mu_{h}\tau_{h}$=1$\times$10$^{-4}$
cm$^{2}$/V. Resolution due to statistical fluctuation of electron-hole
pairs and electronics is not included.  }
\end{figure}

Many methods have been proposed and used to overcome the hole-trapping
problem. One approach is to use the information contained in the pulse
shape\cite{ref:Siffert}. In this approach, the pulses with a slow rise
time are corrected or discriminated by means of specially designed
electronics. A model of the pulse shape is described in
\cite{ref:Bargholtz}. Another approach is the use of a hemispheric
geometry \cite{ref:Siffert}. A logarithmetic field inside the detector
enhances the collection of fast-moving electrons and restrain the hole
collection.  The application of hemispheric CdTe and CdZnTe detectors
to safeguards measurements are described in \cite{ref:Arlt}

\section{High resolution CdTe diode and its application to the stack
detector}

The theoretical energy resolution of CdTe can be calculated from
statistical fluctuations in the number of electron-hole pairs and the
Fano factor ($F$)\cite{ref:Knoll}. By using $\epsilon$ = 4.5 eV and
$F$ = 0.15, the theoretical limit (FWHM) is 200 eV at 10 keV, 610 eV at
100 keV, and 1.5 keV at 600 keV \cite{ref:IEEE2},
 if we could neglect electronic noise.
 These resolutions are very attractive for applications
in astrophysics, where precise determinations of the central energy
and the profile of X-ray and $\gamma$-ray lines are
crucial\cite{ref:NIMA}. For obtaining the ultimate energy resolution
from the CdTe and CdZnTe detectors, collecting full information given
by the transit of both electrons and holes is important. Fig.~7~ is
the result of simulations showing that high bias voltage is
important. In order to obtain a FWHM of 700 eV at the 122 keV line
from $^{57}$Co, a bias voltage of 1400 V should be applied for a
detector with a thickness of 0.5 mm for $\mu_{h} \tau_{h}$ = 1 $\times
10^{-4}$ cm$^{2}$/V.  The thin CdTe device has an advantage over the
thick one because sufficient bias voltage for full charge collection
can be easily applied. Detectors with very low leakage current that
allow such a high bias voltage have been developed by several groups
thorough the use of diode structure either by a blocking electrode or
PIN structure\cite{ref:NIMA,ref:Arlt2,ref:Niraula}.

Takahashi et al. reported a significant improvement in the spectral
properties of CdTe detectors based on the advances made in the
production of high quality CdTe single crystals
\cite{ref:NIMA,ref:SPIE,ref:IEEE}.  The basic idea is to utilize
indium as the anode electrode on the Te-face of the p-type CdTe wafer
with (1,1,1) orientation\cite{ref:Ozaki}. A high Schottky  barrier
formed on the In/p-CdTe interface leads us to the
operation of the detector as a  diode (CdTe diode). The
leakage current of the 2 mm $\times$ 2 mm $\times$ 0.5 mm detector
was 0.7 nA with a bias voltage of 400 V at 20 \degree (Fig. 4).

 Fig.~8~ shows the energy spectrum of $\gamma$-rays from $^{241}$Am
obtained with the CdTe diode.  A bias voltage of 1400 V (internal
electric field of $E$=28 kV/cm) was applied and the operating
temperature was $-$40~\degree. The FWHM of the 59.5 keV line is 0.83
keV corresponding to an energy resolution ($\Delta E/E$) of 1.4\%.
 This is close to the energy resolution of HP-Ge detectors cooled
to liquid nitrogen temperature.  The reduction of the low energy tail
even in the 662 keV line from $^{137}$Cs results in a resolution of
2.1 keV (0.3\%), which is consistent with the theoretical
prediction\cite{ref:IEEE2}. The improvement of the energy resolution by
adopting the Schottky junction is drastic, but the gain and resolution
degrade with time during room temperature operation. It is reported
that both a high electric field of several kV cm$^{-1}$ and a low
operating temperature (below several \degree) ensure stability on time
scales longer than weeks\cite{ref:SPIE}.

\begin{figure}
\label{figure:Am241}
\vspace{1mm}
\centerline{\includegraphics[width=3.3in,clip]{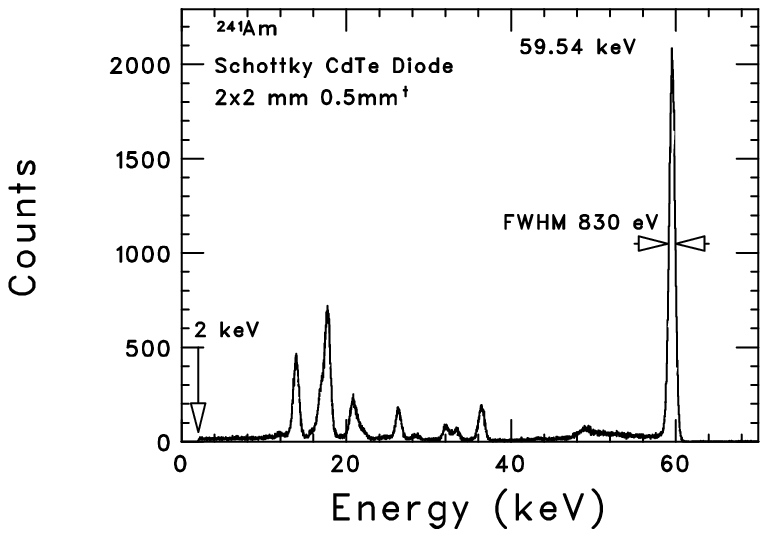}}
\caption[]{$^{241}$Am spectra obtained with the CdTe diode described
in \cite{ref:NIMA}. The energy
resolution at 60 keV is 0.83 keV (FWHM). The detector has a
surface size of 2 mm $\times$ 2 mm and a thickness of 0.5 mm. The time
constant of the shaping amplifier is 2 $\mu$s.  }
\end{figure}

An energy resolution of $\leq$ 1\% at a high photon energy of several
hundred keV under moderate operating conditions is very attractive in
high energy astrophysics. However, good energy resolution with a thick
 CdTe diode will be difficult to achieve as the bias voltage
required for complete charge collection scales with the second power
of the detector thickness.  We, therefore, adopted the idea of a
stacked detector (Fig.~9~), in which several thin CdTe diodes are
stacked together and operated as a single detector.

\begin{figure}
\label{figure:9th}
\vspace{1mm}
\centerline{\rotatebox{-90}{\includegraphics[width=2.7in,clip]{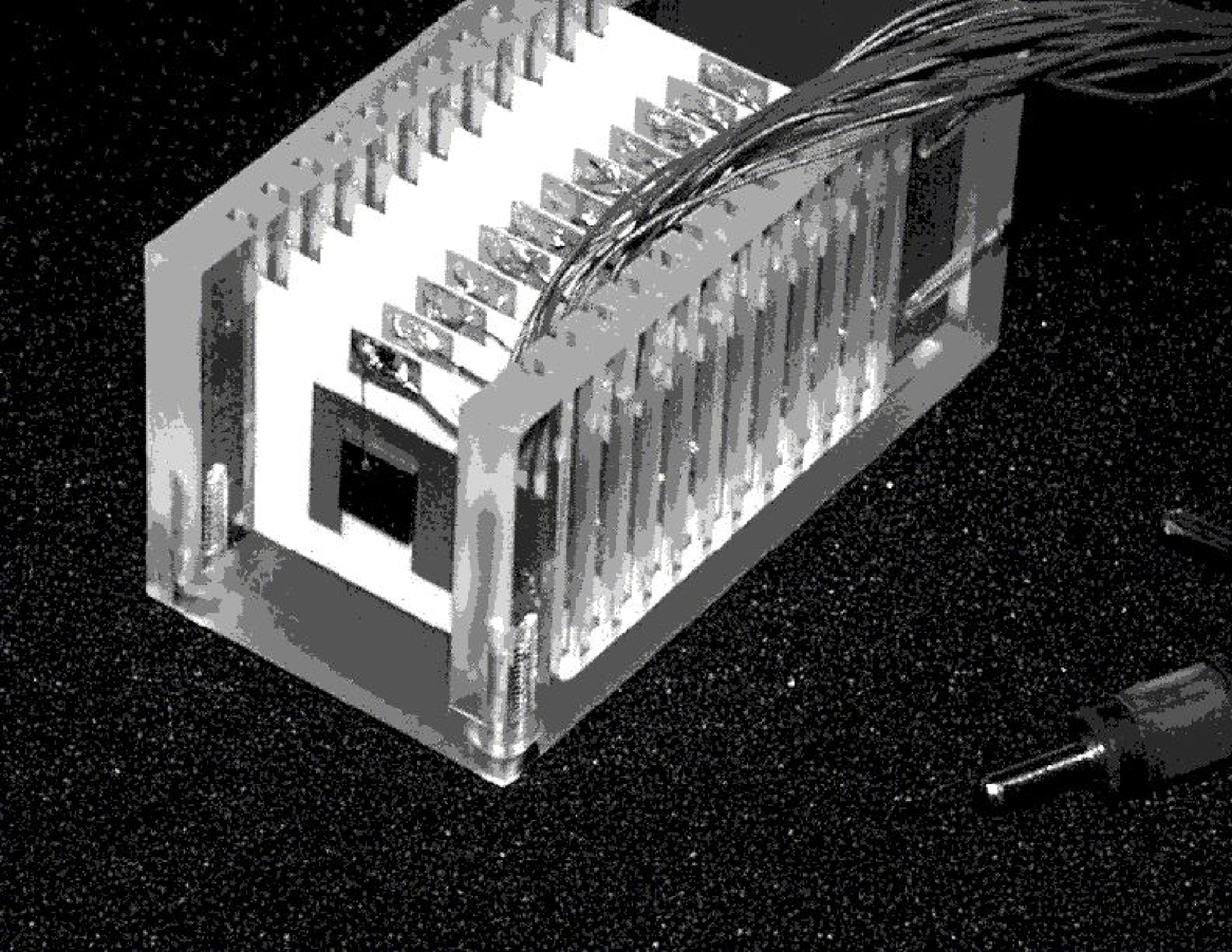}}}
\caption[]{The stacked CdTe diode.  Twelve CdTe diodes with dimentsions
of 5~mm $\times$ 5~mm  $\times$  0.5 mm are stacked together.  The
output from each diode was fed into an individual readout electronics
system, which were operated in anti-coincidence (from \cite{ref:NIMA}.}
\end{figure}

\begin{figure}
\label{figure:10th}
\vspace{1mm}
\centerline{\includegraphics[width=3.15in,clip]{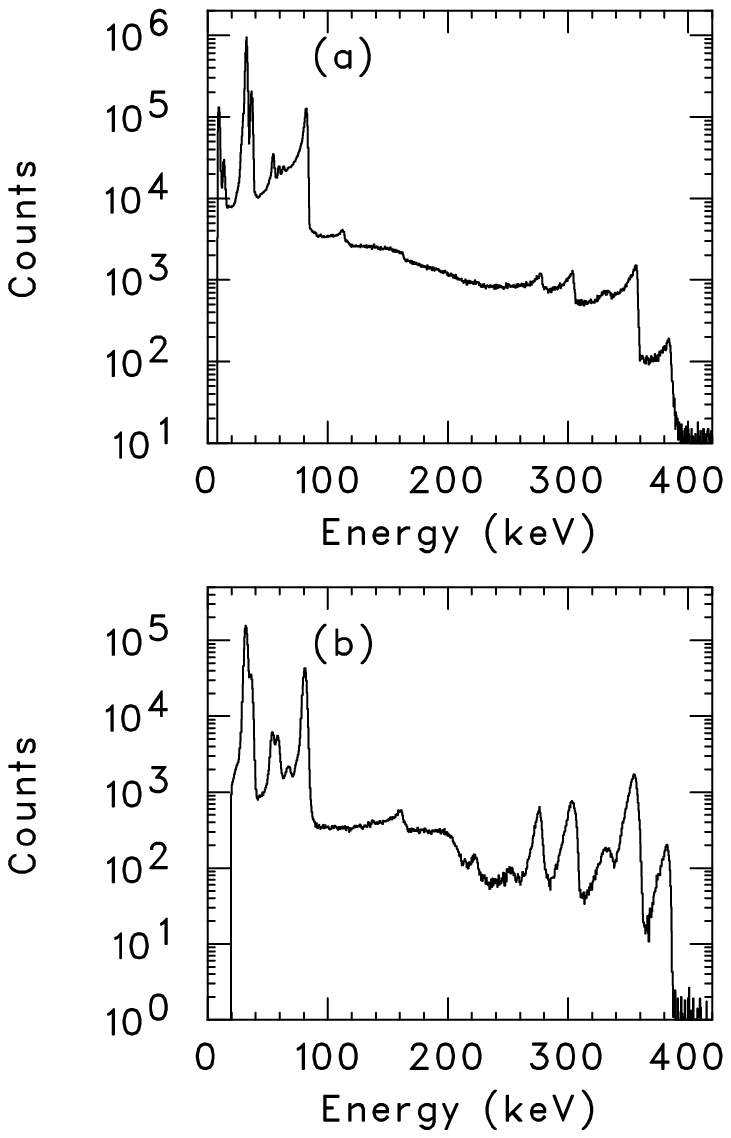}}
\caption{The comparison between the energy spectrum of $^{133}$Ba
$\gamma$-rays obtained (a) from a 2 mm thick planar CdZnTe detector
and (b) from the first four layers of the stack detector with a total
thickness of 2 mm. An asymmetry noticed in the line shape of the
stacked detector above 250 keV is due to the fact that a bias voltage
of 400 V is still not sufficient for full charge collection.}
\end{figure}

Fig.~10~ shows the comparison between the energy spectrum of
$^{133}$Ba $\gamma$-rays obtained from a 2~mm thick planar CdZnTe
detector and that from the first four layers of the stack detector
which consists of 12 layers with a thickness of 0.5 mm, both operated
at 5 \degree. The applied voltages are 300 V for the CdZnTe detector (
$E$ = 1.5 kV/cm ) and 400 V for each layer of the stack detector ( $E$
= 8 kV/cm ). It is clearly shown that the peak to valley ratio at 300
keV is much superior for the stacked CdTe diode as compared to the 2
mm thick CdZnTe detector.  This approach is particularly useful for
energies below 300 keV, where the dominance of photoelectric
attenuation leads to single-site absorption in a single layer. When
$\gamma$-rays with an energy of 300 keV are irradiated onto a 2~mm
thick detector, 9\% of photons deposit all their energy in the
detector. Among them, 75\% are stopped by a single
photoabsorption. For a 6 mm thick detector (12 layers), 44\% and 17\%
of the incident photons are stopped by a single photoabsorption for
200 keV and 300 keV $\gamma$-rays, respectively.

\section{Novel Electrode design for single carrier charge collection}

In order to use CdTe and CdZnTe for the detection of $\gamma$-rays
with energy higher than several hundred keV, the efficiency of
a thick ($>$ 5 mm) detector is necessary. However, application of a
simple planar electrode configuration seems to be difficult, because
the required bias voltage to eliminate low energy tail at the level to
achieve a few keV resolution must be higher than a few hundred
kV. Soon after the emergence of HPB-grown CdZnTe, new ideas based on
the concept of single charge collection have been
proposed\cite{ref:Luke,ref:Barrett,ref:Butler2}. These ideas utilize 
techniques to form strip or pixel electrodes on the surface of the
material and to process many channels by means of high-density analog
LSI in the  form of ASICs. Here we summarize some highlights of the
development of these novel electrode designs.

\subsection {Coplanar grids}
The idea of using the effects of Frisch grid found in gas 
and liquid
detectors was proposed  by Luke in 1994. In this method, an anode
consisting of a  pair of interleaved grids (Coplanar grid) is  formed
on a surface of the detector, as shown in Fig.~11~. A different bias is
applied to each set of electrodes. This potential difference is small 
compared to the overall potential across the detector. When electrons and
holes move within the bulk material of the detector, they induce equal
signals on both grid electrodes. When the electrons come close to the anode
plane, the signal of the grid with the higher potential rises steeply. A net
signal, which is obtained by subtracting the signals from the two grids, is
sensitive to primarily to the   electron signal (single polarity charge
sensing). Since the depth dependent signal due to the movement of holes
is subtracted, the low energy tail in the spectrum can be eliminated. The
dramatic improvement by the coplanar grid is demonstrated in Fig.~12~\cite{ref:Luke2}. 

\begin{figure}
\label{figure:11th}
\vspace{1mm}
\centerline{\includegraphics[width=3.15in,clip]{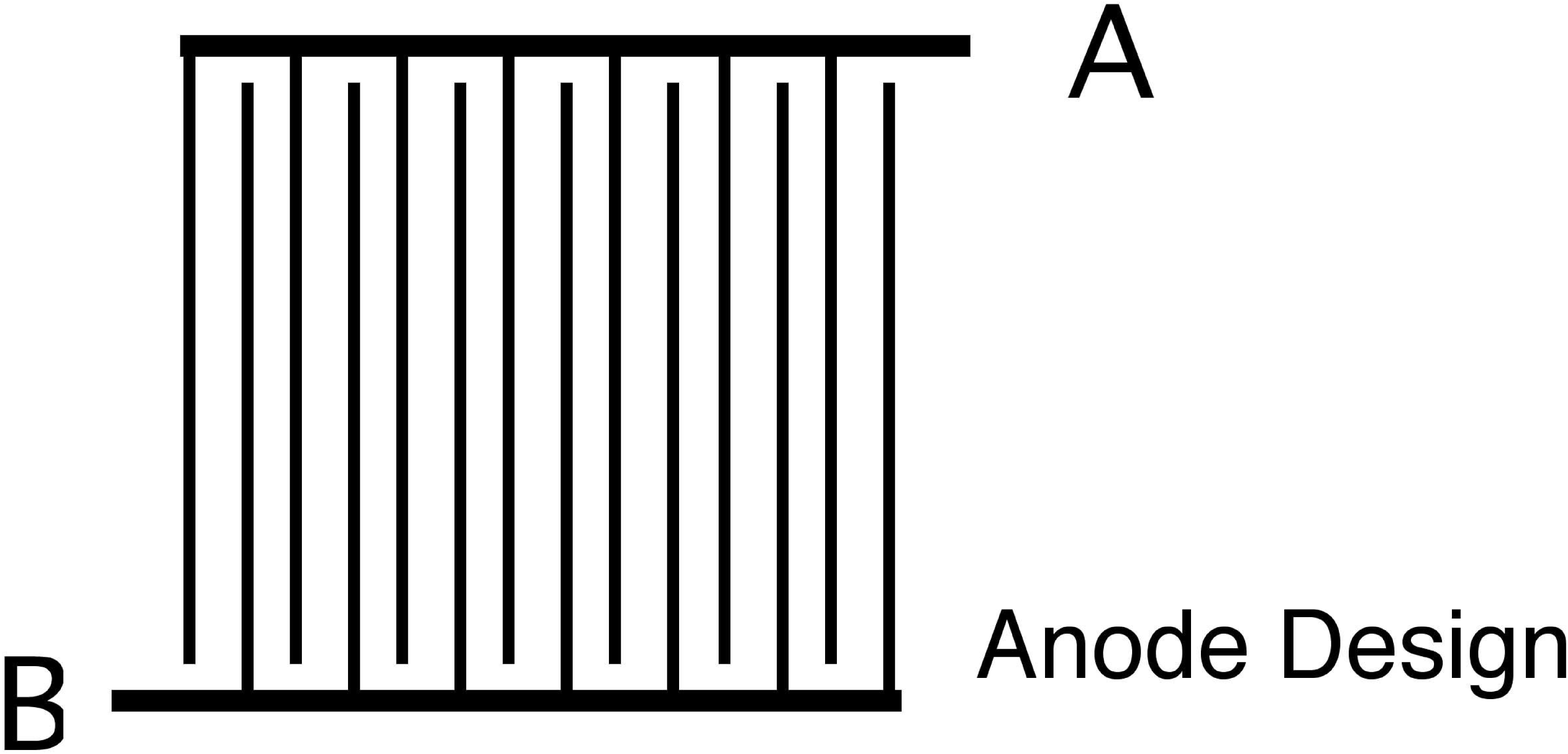}}
\caption{Electrode configuration of coplanar electrodes}
\end{figure}

\begin{figure}
\label{figure:12th}
\vspace{1mm}
\centerline{\includegraphics[width=3.15in,clip]{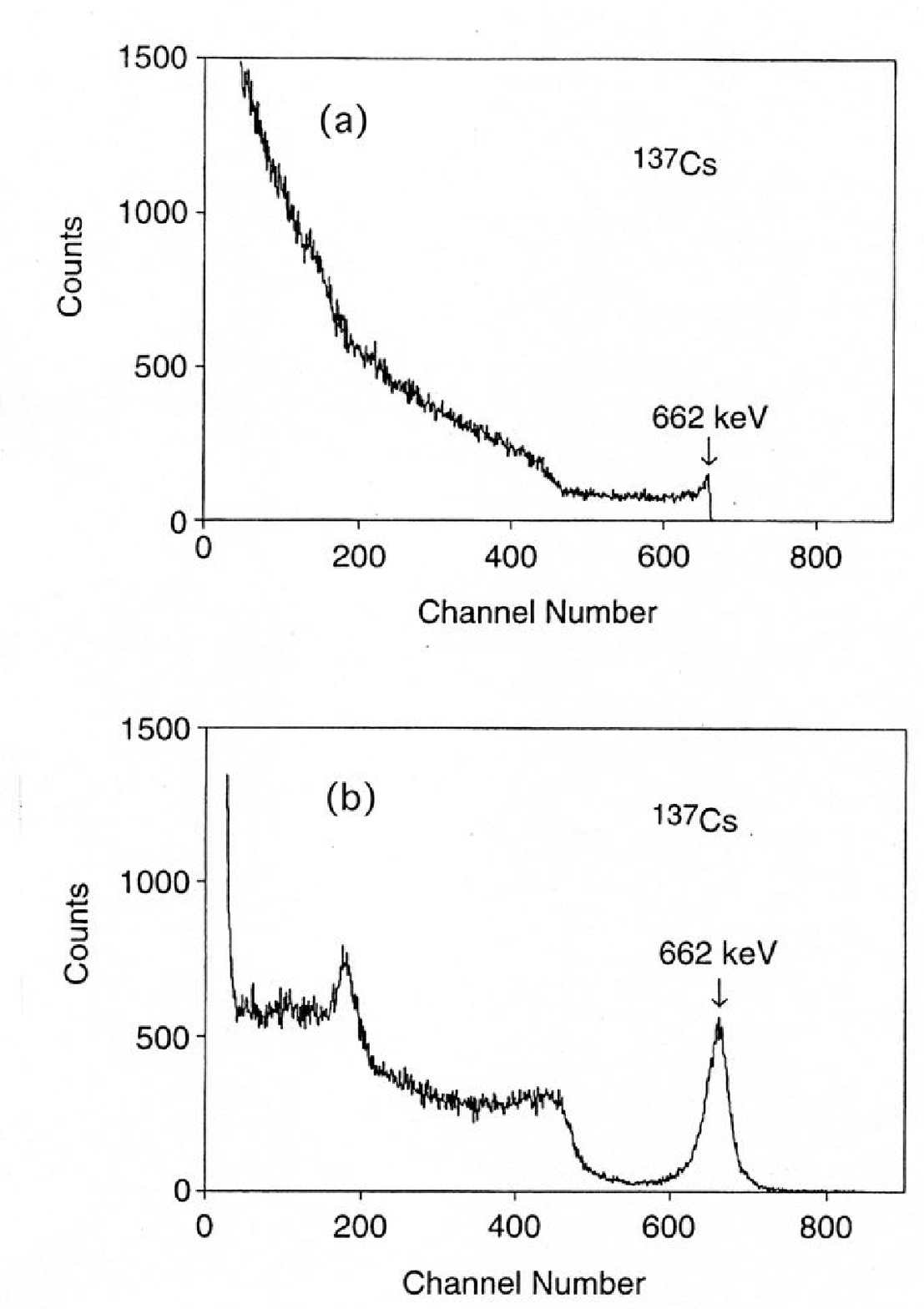}}
\caption[]{Spectra of $^{137}$Cs obtained with the CdZnTe detector (a) in its 
standard configuration and (b) in the coplanar grid configuration.
The dimensions of the detector are 5 $\times$ 5 $\times$ mm$^{3}$.
The applied bias voltage is 500 V and the peaking time of the 
shaping amplifier was set at 2$\mu$s (taken
from
\cite{ref:Luke} )}
\end{figure}

\subsection {Small pixels and control electrode}

It has long been recognized that reducing the size of the electrode
collecting the higher $\mu \tau$ carriers can potentially eliminate
the low-energy tail by effectively achieving single-carrier charge
collection through the "near-field effect". Barret et al have shown
that, if the pixel size is small in comparison with the detector
thickness and the pixel electrode is biased to collect electrons, the
incomplete charge collection due to severe hole trapping can be
dramatically improved, because the induced charge on each pixel anode
is dominated by the number of electrons collected by the
anode\cite{ref:Barrett}.  For nuclear medicine applications,
Barber et al.  applied the idea of ''small pixel effects" to a large
CdZnTe array with a pixellated electrode produced on one side of the
detector slab by photolithography that is bump bonded to a readout
integrated circuit which reads out each pixel
individually\cite{ref:Barber}.  A 64$\times$64 (2.5 cm $\times$ 2.5
cm) CdZnTe arrays with 380 $\mu$m pitch has been constructed and
tested by Barber et al. \cite{ref:Barber-NARA}.  Pixel detectors with
a pixel size of several hundred to a few mm have been developed by
several groups. Optimum spectroscopic performance with electrode
segmentation is discussed in \cite{ref:Shor}.

 Recently, Butler has proposed a new concept based on a control
electrode in addition to a cathode and small
anode\cite{ref:Butler2}. The control electrode surrounds the anode
pixel and accepts a large fraction of the charges induced by the
carriers while they are in transit. Similar to the effect of a
coplanar grid, almost the full charge of the mobile electrons is
induced on the anode.  Mayer et al. extended this concept and
developed an imaging device with orthogonal coplanar anodes
(Fig.~13~)\cite{ref:Mayer}. The signals induced on the strips are used
to determine the interaction position in one dimension. The position
information in the y-direction is inferred from the pixels which are
interconnected to the strips. This electrode configuration provides
information of $n^{2}$ pixels with 2$n$ channels and reduces the
complexity of the readout electronics. The energy resolution (FWHM)
obtained from the detector is 3.42 keV at 60 keV and 5.6 keV at 662
keV.

\subsection {3-D position sensitive semiconductor detector}

In addition to signals from strips or pixels formed on the anode face,
information from the cathode (common electrode) can be used to obtain
the $\gamma$-ray interaction depth. The interaction depth provide a
correction for the electron path length and the trapping which are
important for thick detectors ($\sim$ 1 cm).  He et al. have performed
extensive studies on this possibility\cite{ref:He,ref:He_2}. The fist
3-D position-sensitive semiconductor detector with CdZnTe yielded an
energy resolution of 10 keV at 662 keV and a depth resolution of 0.5
mm. Results from a second generation 12.5$\times$12.5$\times$10
mm$^{3}$ CdTe detector are reported in \cite{ref:Li}.

\begin{figure}
\label{figure:13th}
\vspace{1mm}
\centerline{\includegraphics[width=3.15in,clip]{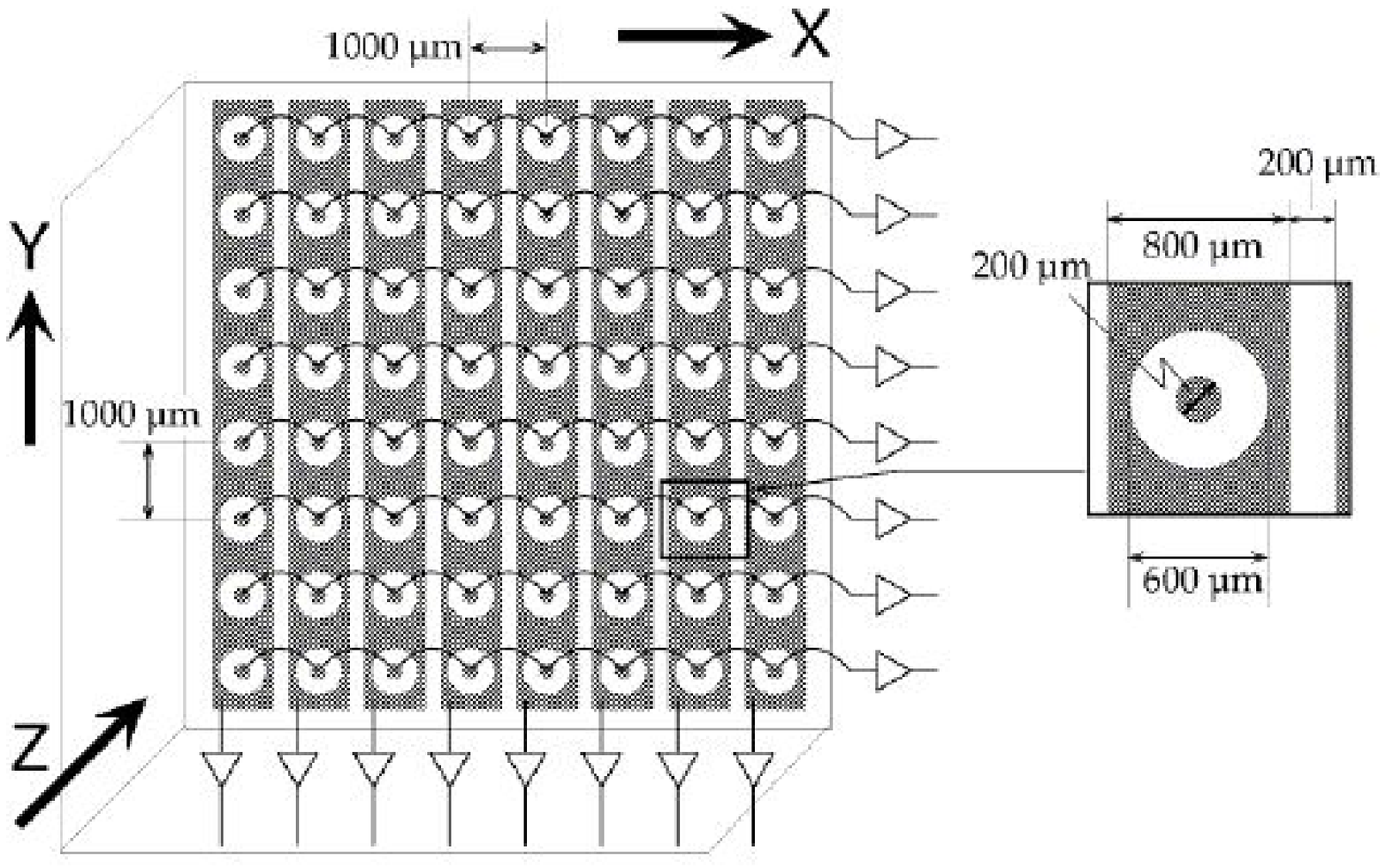}}
\caption[]{The anode geometry of a CdZnTe imaging detector with coplanar
grid and control electrode. The vertical strips measure the $x$-position
while the lines of interconnected pixel provides the $y$-measurement \cite{ref:Mayer} }
\end{figure}

\section{Application to Hard X-ray and $\gamma$-ray mission}

Hard X-rays and $\gamma$-rays are an important frequency window for
expoloring the energetic universe.  The $CGRO$ satellite has revealed
us the fact that a variety of gamma-ray sources exist in the
sky\cite{ref:Gehrels}. However, when compared with X-ray Astronomy,
gamma-ray astronomy is still immature; the sensitivity of instruments
are far from the level achieved by the current X-ray missions
employing focusing telescopes in the energy band below 10 keV.
Gamma-ray instruments in the 21st century should provide much improved
angular and spectral resolution over the instruments in use today.
Cadmium Telluride (CdTe) and Cadmium Zinc Telluride (CdZnTe) solid
state detectors have several promising features which make them
instruments for use as a focal plane imager of a multi-layer grazing
incidence mirror or a coded mask aperture for the next generation of
hard X-ray and $\gamma$-ray astronomy satellites.

\subsection{INTEGRAL}

The IBIS detector on board ESA's INTEGRAL will be the first space
instrument that utilizes the  good spectral resolution of CdTe. Major 
objectives of  the mission are the detection and precise
identification of $\gamma$-ray line spectra to study 
the high energy processes and resulting nuclear interactions
taking place at astrophysical sites. The detector
consists of two layers of
pixellated detector planes  separated $\sim$10 cm and operated
as a focal plane detector of a  coded mask aperture. 
The top layer detector plane is made of
16384 square CdTe detectors (4$\times$4$\times$2 mm$^{3}$)
, which gives a total
sensitive area of 2621 cm$^{2}$. The detector is divided into eight 
rectangular modules of 128 polycells, each polycell containing 16
detector pixels.  In order to improve the energy resolution of the CdTe
detector, a pulse height correction scheme is implemented by a
specially designed ASIC.
According to the measurement of a prototype
detector for IBIS\cite{ref:Limousin}, spectral resolution 
of around 4.5 keV (FWHM) is obtained for the 122 keV line.
The average mobilities are measured to be 946$\pm$50
cm$^{2}$V$^{-1}$s$^{-1}$ for electrons and 79.5$\pm$9
cm$^{2}$V$^{-1}$s$^{-1}$ for holes while the  average lifetime is
reported to be 1.2 $\pm$ 0.1 $\mu$s for  electrons and 
4.6 $\pm$ 0.2 $\mu$s for holes.

\subsection{Swift}

Swift is a first of its kind multiwavelength transient observatory for
$\gamma$-ray burst (GRB) astronomy, due to be launched in 2003. The
BAT instrument on Swift is designed to provide the critical GRB
trigger and quickly measure the burst position to better than 4
arcmin\cite{ref:Swift1,ref:Swift2}.  Since the energy emission from
GRBs peaks at a few 100 keV, the BAT utilizes 32768 CdZnTe detector
with dimensions of 4$\times$4$\times$2 mm$^{3}$ to form a 1.2 m
$\times$ 0.6 m sensitive area in the detector plane. There are 256
detector module (DM) in the BAT. One DM consists of 8$\times$16 array
of CdZnTe elements, which are connected to a 128 channel readout ASIC
(XA chip \cite{ref:IDEAS}).

\begin{figure}
\label{figure:14th}
\vspace{1mm}
\centerline{\includegraphics[width=3.15in,clip]{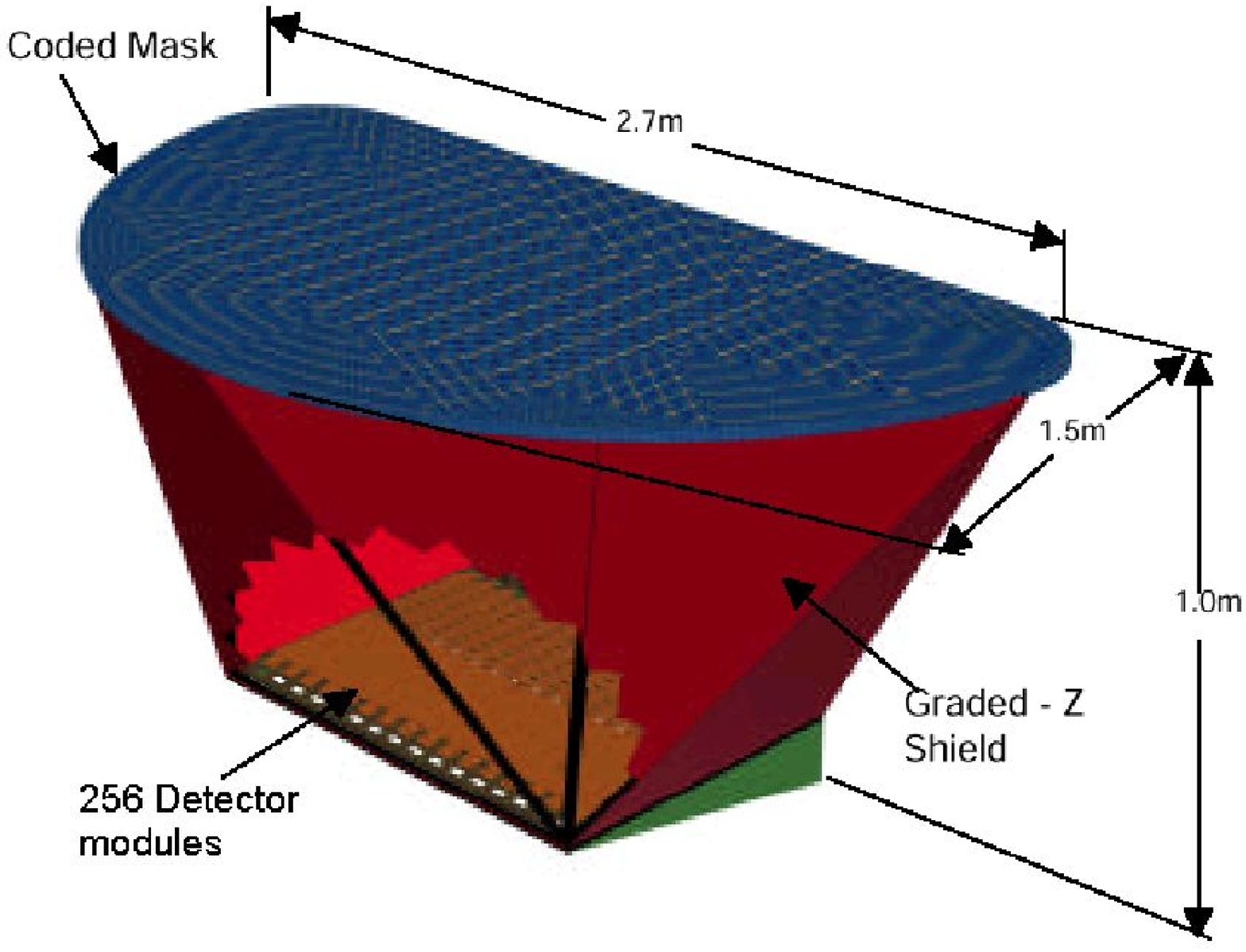}}
\caption[]{Cut away drawing of the BAT instrument onboard Swift
satellite. The D-shaped coded aperture mask is 3 m$^{2}$ with
5 mm pixels. The CdZnTe array consists of 32768 CdZnTe detectors
with dimensitons of 4 mm$\times$4 mm$\times$2 mm. 
Graded-Z shielding reduces the background due to cosmic diffuse 
emission\cite{ref:Swift1,ref:Swift2}}
\end{figure}

\begin{figure}
\label{figure:15th}
\vspace{1mm}
\centerline{\includegraphics[width=3.00in,clip]{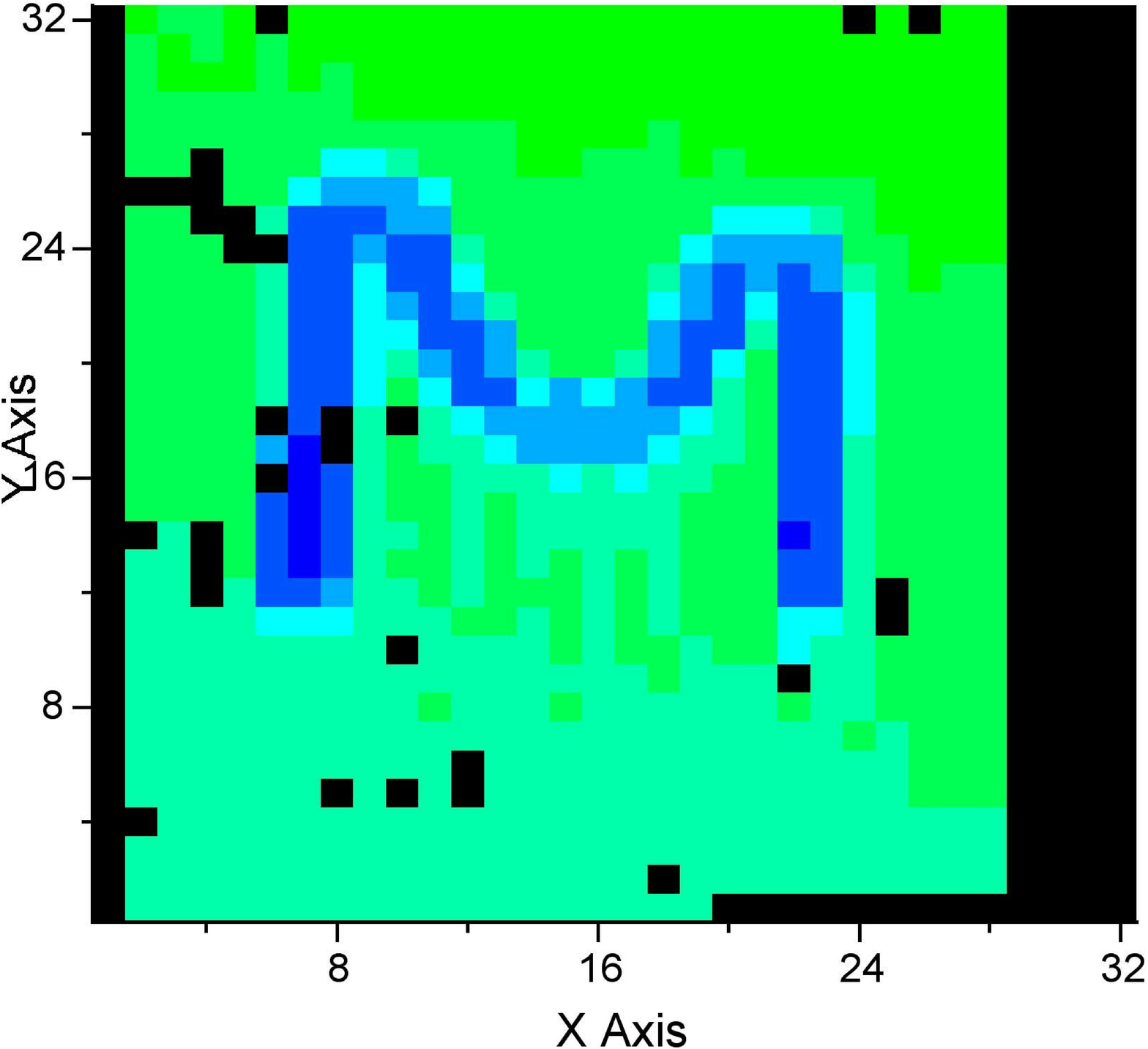}}
\caption{Radiographic image of an object made of soldering wire 
irradiated by $\gamma$-rays from $^{241}$Am  
obtained
with the  32$\times$32 CdTe pixel detector covering 6.8 mm $\times$ 6.8
mm at room temperature. The pixel size is 200 $\mu$m $\times$ 200 $\mu$m.}
\end{figure}

\subsection{Future Approach}

For future X-ray astronomy missions, one of the main objectives is
observation with a very high sensitivity in the 10-100 keV band,
where non-thermal emission becomes dominant over thermal
emission\cite{ref:NIMA}. This could be achieved by employing a
multi-layer, grazing incidence hard X-ray telescope (``super mirror'')
in conjunction with a hard X-ray imaging detector as a focal plane
detector.  The current goal for the imaging detector is a CdTe or
CdZnTe pixel detector with both a fine position resolution of a few
100 $\mu$m and a high energy resolution better than 1 keV (FWHM) in
this energy range.
In order to cover the field of view of the
telescope, the detector should have an area of several cm$^{2}$.  Fast
timing of several hundred ns will be required for the active
shielding, which would be necessary to achieve a low background
environment in space. 

 Research and development projects toward these goals are now under
way by several
groups\cite{ref:Stahle,ref:Matteson,ref:IEEE2,ref:Sushkov}. To
realize fine pitch (finer than several hundred microns) CdTe and/or
CdZnTe pixel detectors, good detector material and low noise
amplifiers and a readout system for more than 10,000 independent
channels will be the key technology. A simple and robust connection
technology needs to be established, because high compression and/or
high ambient temperature would damage the CdTe and CdZnTe
crystal. Fig.~15 shows the radiograpic image obtained with fine pixel
detectors developed under the collaboration between Bonn Universty and
ISAS\cite{ref:Fischer} . The size of the pixels is 200$\mu$m $\times$
200 $\mu$m. They are directly bump bonded to a two-dimensional photon
counting ASIC (MPEC2) by using newly developed gold-stud bump bonding
technology\cite{ref:IEEE2}.

In addition to pixel detectors, a large area
array ($\sim$ 5000 cm$^{2}$) consisting of strips has been reported
in \cite{ref:Stahle2}. The detector is a 6$\times$6 array of CdZnTe
strip detectors. Orthogonal patterns are imprinted on the
top and bottom sides of each detector. With strips separated by
100 $\mu$m, a spatial resolution of $<$ 50 $\mu$m has been demonstrated.

\section{Conclusion}

Owing to the significant progress in producing high quality CdZnTe and
CdTe crystals, these materials are now regarded as ``serious''
candidates for the next generation of hard X-ray and $\gamma$-ray
detectors. Applications to nuclear medicine are reviewd in
\cite{ref:Barber-NARA,ref:Scheiber,ref:Eisen2}. Good energy resolution better than 1
keV is achieved for 60 keV $\gamma$-rays by the diode structure for a
thin detector. A novel electrode configuration solves the problem of
incomplete charge collection for the detection of high energy
$\gamma$-rays with a thick detector. As discussed by Fougeres et
al. \cite{ref:Fougeres}, the choice between CdTe and CdZnTe is still
very difficult to make. There is still some room for improvements for
real applications. These include the production of large and uniform
single crystals, especially for the HPB-grown CdZnTe, and the bump
technology specific to ``fragile'' CdTe and CdZnTe. The cost of the
material is still high for a large scale $\gamma$-camera.  Further
advances in application-specific integrated circuits (ASICs) are
necessary for the fabrication of fine-pitch pixel detectors.
Extensive studies currently underway by many laboratories will boost
the further advances of of these interesting materials .

\section{Acknowledgment}
We acknowledge constructive discussion with  R. Ohno and C. Szeles. 
We thank G. Sato, M. Kouda, Y. Okada for the help of experiments and
M.D. Audley for his critical reading of the manuscript.

\end{document}